\documentstyle[11pt]{article}
\begin{document}

\newcommand{\be}{\begin{equation}}
\newcommand{\ee}{\end{equation}}
\newcommand{\bea}{\begin{eqnarray}}
\newcommand{\eea}{\end{eqnarray}}
\newcommand{\bean}{\begin{eqnarray*}}
\newcommand{\eean}{\end{eqnarray*}}
\newcommand{\bear}{\begin{array}}
\newcommand{\eear}{\end{array}}
\newcommand{\half}{\frac{1}{2}}
\newcommand{\Tr}{{\rm tr}\,}
\newcommand{\raw}{\rightarrow}
\newcommand{\law}{\leftarrow}
\newcommand{\lraw}{\leftrightarrow}
\newcommand{\inp}[2]{\langle{#1}|{#2}\rangle}
\newcommand{\bra}[1]{\langle{#1}|}
\newcommand{\ket}[1]{|{#1}\rangle}
\newcommand{\eps}{\varepsilon}
\newcommand{\dint}{\int\!\!\!\int}
\newcommand{\im}{\imath}
\newcommand{\es}{\,^S\!E^a_i}
\renewcommand{\thefootnote}{\fnsymbol{footnote}}

\title{\sc The weaving of curved geometries}
\author{\sc Joost Zegwaard\thanks{e-mail: zegwaard@fys.ruu.nl} \\ \\ \it
Instituut voor Theoretische Fysica \\ \it Postbus 80.006 \\ \it 3508 TA Utrecht
\\ \it The Netherlands}
\date{}
\maketitle

\begin{abstract}
In the physical interpretation of states in non-perturbative loop quantum
gravity the so-called weave states play an important role. Until now only
weaves representing flat geometries have been introduced explicitly. In this
paper the construction of weaves for non-flat geometries is described; in
particular, weaves representing the Schwarzschild solution are constructed.
\end{abstract}

\vspace{-12.5cm}

\begin{flushright}
 October, 1992 \\ THU-92/28 \\ hepth@xxx/9210033
\end{flushright}

\vspace{11cm}

\section{Introduction}

The new variables for canonical gravity, discovered by Ashtekar \cite{canon},
and the subsequent progress in the formulation of a quantum theory based on
these variables \cite{jacsmo,rovsmo} have caused renewed interest in quantum
gravity from the canonical point of view. In particular the so-called loop
representation \cite{rovsmo,review} is interesting, because in this
representation a set of solutions for canonical quantum gravity is known
explicitly. These solutions are essentially the knot and link invariants of
sets of smooth, non-intersecting loops. The interpretation of these solutions
however is obscure: in the
construction of the loop representation the physical interpretation has got
lost; moreover, the correct inner product is still unknown. In order to gain
insight in the physics of the loop representation comparable procedures have
been applied to better understood theories like the quantized Maxwell field
\cite{maxwell} and linearized gravity \cite{gravton}. A procedure has been
proposed to encapture the states and operators of linearized gravity into the
loop representation of the full quantum theory, which sheds some light on the
physical meaning of the states in the latter \cite{review,zeg2}.

Weaves \cite{nonper,thread} play an essential role in the interpretation of
physical states. Their role in the theory is twofold. Firstly, they represent
in a well-defined way a smooth classical geometry on large scales, while
exhibiting a discrete structure at the Planck scale. Secondly, perturbations on
these weaves (more precisely: the way physical functionals vary on weave
perturbations) may represent graviton-like perturbations of the classical
geometry. In this letter I will particularly emphasize on the first aspect of
weaves; for the second aspect I refer to other papers \cite{zeg2,rovelli}.

\section{Weaves for flat space}

In the Ashtekar formalism the metric on the spatial manifold $\Sigma$ is
defined in terms of the inverted densitized triad $E^a_i$:
\be
 Q^{ab} \equiv (\det q) q^{ab} = E^a_i E^{b,i} ,
\ee
where $q_{ab}$ is the three metric on $\Sigma$, $q^{ab}$ its inverse. For any
smooth 1-form $\omega$ on $\Sigma$ one can define the functional
\be
 Q(\omega) = \int_\Sigma d^3x\, \sqrt{E^a_i(x) E^{b,i}(x)
  \omega_a(x) \omega_b(x)}  \label{classQ}.
\ee
In terms of the $T$-algebra of ref.\ \cite{rovsmo} it can be expressed as
\be
 Q(\omega) = \lim_{\eps\raw 0} \int_\Sigma d^3x \left(\int d^3y\int
  d^3y'\,f_\eps(x-y)
  f_\eps(x-y') (-\frac{1}{4})T^{ab}[\gamma_{yy'}](0,\half) \omega_a
  (y) \omega_b(y')\right)^\half ,
\ee
where $\gamma_{yy'}$ is a smooth loop such that $\gamma_{yy'}(0)=y$,
$\gamma_{yy'}(\half)=y'$, shrinking to the point $y$ in the limit $\eps\raw 0$;
$f_\eps$ is a smearing function which goes to a delta function in this limit.
After quantization this leads to the following expression (in bra/ket notation
of ref.\ \cite{thread}):
\be
 \bra{\alpha}\circ \hat{Q}(\omega) = \frac{\hbar G}{2}
  \left( \lim_{\eps\raw 0}
  \int_\Sigma d^3x\,\sqrt{F^a_\eps(x,\alpha) F^b_\eps(x,\alpha) \omega_a(x)
  \omega_b(x)} \right) \bra{\alpha}\ , \label{defQ}
\ee
where $F^a_\eps(x,\alpha) = \int ds\, f_\eps(x-\alpha(s)) \dot{\alpha}^a(s)$.
A weave for a metric $\bar{Q}$ is defined as a set of loops such that the
prefactor on the right hand side equals $\bar{Q}(\omega)$, as defined above.
This set of loops is in general defined in terms of a flat background metric;
the final result is not allowed to depend on this auxiliary metric anymore. For
instance, the distances between the loops as measured in the resulting metric
do not depend on the auxiliary metric.

The original idea for the construction of the weave was based on a
regularization using the triad structure \cite{review}. The weave must,
according to this proposal, be split into three parts $W_i$ ($i=1 \ldots 3$).
If $\bar{Q}^{ab}$ is  a classical metric constructed from a triad
$\bar{E}^a_i$, the corresponding weave `components' $W_i$ should satisfy
\be
 \frac{\hbar G}{2} \lim_{\eps\raw 0} \int_\Sigma d^3x\, F^a_\eps(x,W_i)
  \omega_a(x) = \frac{\hbar G}{2} \oint_{W_i}\omega =  \int_\Sigma d^3x\,
  \bar{E}^a_i(x) \omega_a(x)  \label{reg}
\ee
for 1-forms $\omega$ which are slowly varying on the Planck scale. The reason
for the requirement for $\omega$ to be slowly varying is that the weave will
typically exhibit a discrete structure on the Planck scale, therefore the
expression on the left hand side in eqn.\ \ref{reg} will not see variations of
$\omega$ on this scale. The simplest example of a weave satisfying eqn.\
\ref{reg} is constructed for the flat geometry, $\bar{Q}^{ab} = \delta^{ab}$
and $\bar{E}^a_i = \delta^a_i$. The equation
\be
 \frac{\hbar G}{2}\int ds\, \dot{W}^a_i(s) \omega_a(W_i(s)) = \int_\Sigma
  d^3x\, \delta^a_i \omega_a(x)
\ee
is fulfilled if $W_i$ consists of lines $l_i$ in the $\hat{e}_i$-direction, of
which the intersection with the $x_i=0$ plane is a square lattice with lattice
distance $d=\sqrt{\half\hbar G}= l_p\sqrt{\half}$, where $l_p$ is the Planck
length (in units such that $c=1$). Note that this weave is not unique at all:
slowly varying one-forms $\omega$ will not see perturbations of the weave on
small scales, so every weave differing from this one only on small scales will
satisfy eqn.\ \ref{reg}; in particular one is allowed to add small `threads' to
close the lines into loops again. For convenience I will further on refer to
this type of weave as `large-loop weave'.

Using the above weave it appeared to be relatively simple to describe graviton
states, which were constructed in a loop representation for the linearized
theory in ref.\ \cite{gravton} in terms of exact loop functionals
\cite{review,zeg2}. However, an important objection against this weave is that
it is not
isotropic: there are three preferred directions. One of the consequences is
that the area operator as defined in refs. \cite{nonper,thread} does not give
the correct value for this weave. The reason is probably that there is no
theoretical justification for the splitting of the weave into the three
components.

Recently another proposal has been made for constructing weaves, which avoids
the choice of a triad and which is isotropic (for a flat geometry) on large
scales \cite{thread}. These weaves essentially consist of a large number of
small loops (order of Planck length) distributed over the spatial manifold.
I will call this type of weave the `small-loop weaves'. The density and
orientation of the loops determine the precise expression for the metric via
the relation
\be
 \half \hbar G \oint_\Delta |\omega| = \int_\Sigma d^3x\, \bar{Q}(\omega) ,
\ee
where $\omega$ is again a slowly varying smooth 1-form and $\Delta$ is the
weave. The left hand side of this equation is calculated by evaluating eqn.\
\ref{defQ} for a set of smooth, non-intersecting loops. From this relation a
weave for the flat metric can be checked to be a set of Planck-sized loops,
distributed with random orientation and uniform density of four Planck lengths
of curve per Planck volume over the spatial
manifold. If the loops are chosen to be circles with radius $a$ and the density
of circles equals $n=a^{-3}$, then $a$ equals $l_p\sqrt{\pi/2}$ \cite{thread}.
It can be shown that the area operator, measuring the area of a surface by
counting its number of intersections with the weave and multiplying by $\half
l_p\!^2$, reproduces for this weave the classical results on large scales. The
description of graviton waves in this new picture is more complicated, but some
first results seem to confirm that it is possible to incorporate them
\cite{rovelli}.

Although the precise definition of the two types of weaves as described above
is different, the basic idea is the same: a weave is a dense structure of
lines, distributed over the spatial manifold with a typical discreteness of the
order of the Planck length.

\section{Schwarzschild geometry with large loops}

In the previous section weaves representing flat geometries were described. One
expects that weaves for curved geometries can be constructed in a similar way.
In this section a large-loop weave corresponding to the spatial part of the
Schwarzschild geometry will be constructed; in the next one we will consider
small-loop weaves for more general curved spaces and the Schwarzschild geometry
in particular. Since weaves seem to be suited only for approximating positive
metrics, the Schwarzschild geometry will only be considered outside its
horizon.

Suppose $\es$ is a triad field generating the spatial Schwarzschild geometry
$q^S_{ab}$. The weave $N=N_1\cup N_2\cup N_3$ representing this triad should
obey the following equation:
\be
 \frac{\hbar G}{2} \oint_{N_i} \omega =
  \int_\Sigma d^3x\,\omega_a(x)\es(x)
\ee
for slowly varying $\omega$.

In spherical coordinates the spatial Schwarzschild metric reads:
\be
 ds^2 = \left(1 - \frac{2m}{r}\right)^{-1} dr^2 + r^2 d\Omega^2 .
\ee
A straightforward calculation shows that a possible triad for this metric in a
Cartesian coordinate system is
\be
 \es = \delta^a_i + e^a_i ,
\ee
where
\be
 e^a_i = \epsilon^a_{\;ik} \frac{x^k}{r}\sqrt{\frac{2m}{r-2m}}
\ee
($r=\sqrt{\sum x_i^{\,2}}$).

Having constructed a triad corresponding to the Schwarzschild geometry one has
to find a weave generating it. The triad can be split into a flat symmetric
part $\delta^a_i$ and an antisymmetric part $e^a_i$. The flat part will be
represented by the flat weave $W$ (but only outside the sphere $r=2m$); another
weave $M$ representing $e^a_i$ will be constructed. The two weaves together
then generate the full spatial Schwarzschild geometry.  For definiteness take
$i=1$. Since
\be
 \vec{e}_1 = \frac{1}{r}\sqrt{\frac{2m}{r-2m}}\left(
 \bear{c}
   0 \\ -z \\ y
 \eear \right) ,
\ee
the following equation should hold:
\be
 \frac{\hbar G}{2}\oint_{M_1} \omega =
  \int_\Sigma d^3x\, \omega_a(x) e^a_1(x)  \label{basis} .
\ee

The expression for $\vec{e}_1$ suggests that loops in $M_1$ must be chosen to
be circles centered in and perpendicular to the $x_1$-axis. Loops $l_1$ in
$M_1$ are characterized by two integers $n_1$, $n_2$:
\be
 l_1^{n_1n_2}(s) = (n_1 d, r(n_1,n_2)\cos 2\pi s, r(n_1,n_2)
  \sin 2\pi s)
\ee
(again $d=\sqrt{\half\hbar G}$); $l_2$ and $l_3$ are defined by cyclic
permutations of the components. The circle radius is defined by
\be
 r(n_1,n_2) = \sqrt{R(n_2)^2 - (n_1 d)^2} ,
\ee
where
\be
 R(n_2) = 2m + \frac{d^2 n_2\!^2}{8m} ,
\ee
which means that the loops with a fixed value of $n_2$ are situated on a sphere
with radius $R(n_2)$ around the origin. Note further that $n_2 \geq 0$,
$|n_1|\, d \leq R(n_2)$.

To check that $M$ indeed generates the Schwarzschild metric, we must integrate
a 1-form $\omega$ over, for instance, $M_1$:
\bea
 \lefteqn{\frac{\hbar G}{2} \oint_{M_1} \omega =
  \frac{\hbar G}{2}\sum_{n_1,n_2} \int_0^1 ds\, \omega_a(l_1^{n_1n_2}(s))
  (\dot{l}_1^{n_1n_2})^a(s)=} \nonumber \\
 &&\frac{\hbar G}{2}\sum_{n_1,n_2} \int_0^1 ds\, \omega_a(n_1 d,
  r(n_1,n_2)\cos 2 \pi s,  r(n_1,n_2)\sin 2 \pi s) \nonumber \\
 &&\times\ 2\pi r(n_1,n_2) \left(
 \bear{c}
  0 \\ -\sin 2 \pi s \\ \cos 2 \pi s
 \eear \right)^a. \label{explicit}
\eea
Replace discrete coordinates $n_k$ by continuous ones: $x=n_1 d$, $\phi = 2 \pi
s$, $r= r(n_1,n_2)$, $y=r \cos \phi$, $z = r \sin \phi$ to find that this is
approximately equal to
\be
 \frac{\hbar G}{2} \oint_{M_1} \omega =
  \int_\Sigma d^3x \,\omega_a(x) \frac{1}{r}\sqrt{\frac{2m}{r-2m}} \left(
  \bear{c}
   0 \\ -z \\ y
  \eear \right)^a = \int_\Sigma d^3x \,\omega_a(x) e_1^a(x) .
\ee
Putting $W$ and $M$ together gives a new weave $N= W\cup M$, generating the
full Schwarz\-schild geometry outside $r=2m$. It is clear that at large
distances the influence of $M$ becomes negligible, while close to the horizon
the majority of lines belongs to $M$.

\section{Curved geometries with small loops}

As was noted in ref.\ \cite{thread} it is not very difficult to deform a flat
weave into a weave for a curved geometry close to the flat one. Suppose that
the discrepancy between a flat metric $h^{ab}$ and a curved one $Q^{ab}$ is
described by a tensor field $t$:
\be
 Q^{ab} = t^a\!_c t^b\!_d h^{cd}.
\ee
Suppose further that the flat metric $h$ is approximated by a weave $\Delta$.
Using $t$ a weave $\Delta_t$ generating $Q$ can easily be constructed. Consider
a typical
loop contained in $\Delta$, a small circle around a point $x$ denoted by $x +
\gamma$. A deformed circle is created by the action of $t$ on $\gamma$:
\be
 x^a + \gamma^a(s) \raw x^a + t^a\!_b(x) \gamma^b(s) .
\ee
The tangent vector to the circle in $s$ becomes $t^a\!_b\dot{\gamma}^b(s)$,
while the density of circles --- the number of circles per unit volume --- is
constant. When $\omega$ is integrated over the deformed weave $\Delta_t$, one
gets
\be
 \int_{\Delta_t}|\omega| =
  \int_\Sigma d^3x\, \sqrt{h^{cd} (t^a\!_c \omega_a) (t^b\!_d\omega_b)}
  = \int_\Sigma d^3x\, Q(\omega) .
\ee
This construction however fails for metrics which are not slowly varying or not
bounded, for instance the Schwarzschild solution near the horizon. The problem
is that the original circles may be deformed into macroscopical loops. This
problem can be circumvented by increasing the density of loops, such that the
average length of the loops stays of the order of the Planck length.

As a specific case the weave for the Schwarzschild geometry can be worked out.
In spherical coordinates $Q_S(\omega)$ can be written as
\be
 Q_S(\omega) = \sqrt{\omega_r^2 + k(r)^2 r^2 \omega_\theta^2 +
  k(r)^2 r^2 \sin^2\!\theta\, \omega_\phi^2},
\ee
where $k(r) = \sqrt{r/(r-2m)}$ and $\omega_r$, $\omega_\theta$ and
$\omega_\phi$ are the spherical components of $\omega$. $Q_S$ can be derived
form the flat metric by means of the tensor field $t$, satisfying
\be
 t^r\!_r =1 \ ; \ \ \ t^\phi\!_\phi = k(r) \ ; \ \ \ t^\theta\!_\theta
  = k(r) \ ;
\ee
all other components are zero. If $\Delta$ is deformed by means of this tensor
field, then the density of loops is held invariant, but the tangential
components $\gamma^\theta$ and $\gamma^\phi$ are expanded by a factor $k(r)$.
These loops therefore become macroscopic close to the horizon. Instead, it is
possible to increase the density of loops at radius $r$ with a
factor $k(r)$, from $n$ to $n k(r)$, leading to an overall factor $k(r)$ in
front of $Q(\omega)$; the effect in the radial direction must then be balanced
by multiplying the radial components $\gamma_r$ by a factor $k(r)^{-1}$. When
integrating a vector field over the thus obtained weave $\Delta_S$ one gets
\bea
 \half l_p^2 \oint_{\Delta_S} |\omega|&=&\int_\Sigma d^3x\, k(r)
  \sqrt{\frac{\omega_r^2}{k(r)^2} + r^2 \omega_\theta^2 +
  r^2 \sin^2\!\theta\, \omega_\phi^2} \nonumber \\
 &=& \int_\Sigma d^3x\, \sqrt{\omega_r^2 + k(r)^2 r^2 \omega_\theta^2 +
  k(r)^2 r^2 \sin^2\!\theta\, \omega_\phi^2}.
\eea
Because of the specific expression for the factor $k(r)$, the density of loops
can go to infinity near the horizon while the total number of loops remains
finite; one can calculate the number of loops between $r=2m$ and $r=2m +
\varepsilon$, where $\varepsilon \ll 2m$ ($A =16 \pi m^2$ is the horizon area):
\bea
 N &=& n \int_{2m}^{2m+\varepsilon} dr\, 4\pi r^2\sqrt{\frac{r}{r-2m}}
  \simeq \half nA\sqrt{2m\eps} .
\eea
Note that for the large-loop weave also a finite extra numbers of lines was
needed to construct the Schwarzschild solution up to the horizon. Furthermore,
in this new picture we also see that the
extra lines are tangential to spheres centered around the origin (the radial
contribution of the loops in this weave stays the same, since the factors
$k(r)$ in the density and $k(r)^{-1}$ in the radial components of the loops
cancel against another). Thus, although the construction differs, the final
qualitative picture is the same for both types of weave: first a sphere with
radius $r=2m$ is cut out, then the density of tangential lines close to the
horizon is increased in the specified way. One can check that the weave we
described also generates the right surface areas outside the horizon.

To calculate the number of intersections with the horizon itself, first
note that the fraction of loops at distance $d$, $d<a^2/2m$, to the horizon,
which intersects this horizon, is $\sqrt{1-2md/a^2}$. It follows
from a simple calculation that the total number of loops intersecting
the horizon is approximately $A/2l_p\!^2$, so the total number of
intersections is $A/l_p\!^2$. As a consequence, given the exterior
structure of the Schwarzschild solution in terms of this weave there is
a countable number of possible internal states, constructed by
reconnecting the loops inside the horizon. Consider two loops
intersecting the horizon. Assume that there are only two physically
independent ways to reconnect them inside, namely linked or non-linked;
multiple linkings and knots are excluded.  This seems to be a
reasonable assumption, since in the weaves as we defined them knotted
loops and multiple links do not appear anyway. It is now easy to see
that the number of inequivalent possible reconnections is
$2^{A/2l_p\!^2}$. This number corresponds to an `entropy' $\half (\log 2)
A/l_p\!^2$, which is in fact quite close to the well-known value of
the physical entropy of a spherically symmetric black hole,
$S=\frac{1}{4}A/l_p\!^2$. Although this calculation of entropy is probably too
simple, the weave picture does give an idea about a possible origin of
entropy in pure quantum gravity in the loop representation.

As a final remark I would like to point out that the weaves as described in
this paper a priori describe only the construction of metrics in the
unconstrained state space, which are therefore not necessarily solutions to
Einstein's equations. It is easy to take the vector constraint into account by
considering link classes of weaves, which generate geometries (metrics modulo
diffeomorphisms). The role of the scalar constraint however, and the related
issue of time and dynamics, require further investigation.


\begin{thebibliography}{10}
 \bibitem{canon}
  A. Ashtekar, Phys. Rev. D 36 (1987) 1587
 \bibitem{jacsmo}
  T. Jacobson and L. Smolin, Nucl. Phys. B 299 (1988) 295
 \bibitem{rovsmo}
  C. Rovelli and L. Smolin, Nucl. Phys. B 331 (1990) 80
 \bibitem{review}
  C. Rovelli, Class. Quant. Grav. 8 (1991) 1613
 \bibitem{maxwell}
  A. Ashtekar and C. Rovelli, Class. Quant. Grav. 9 (1992) 1121
 \bibitem{gravton}
  A. Ashtekar, C. Rovelli and L. Smolin, Phys. Rev. D 44 (1991) 1740
 \bibitem{zeg2}
  J. Zegwaard, Nucl. Phys. B 378 (1992) 288
 \bibitem{nonper}
  L. Smolin, `Recent developments in nonperturbative quantum gravity',
   Syracuse preprint (1992) (hep-th/9202022)
 \bibitem{thread}
  A. Ashtekar, C. Rovelli and L. Smolin, Phys. Rev. Lett. 69 (1992) 237
 \bibitem{rovelli}
  J. Iwasaki and C. Rovelli, `Gravitons as embroidery on the weave',
  Pittsburgh preprint (1992)
\end{thebibliography}
\end{document}